# Implementation of a Directional Modulation Testbed for Reconfigurable Transmitters for Spatially Agile MIMO Systems

Jonathan E. Swindell, *Graduate Student Member, IEEE*, David W. Cox, *Graduate Student Member, IEEE*, Rebekah Edwards, *Graduate Student Member, IEEE*, Emma Lever, *Member, IEEE*, Adam C. Goad, *Member, IEEE*, Austin Egbert, *Member, IEEE*, Charles Baylis, *Senior Member*, *IEEE*, and Robert J. Marks II, *Life Fellow*, *IEEE*

*Abstract*—This paper demonstrates the implementation and validation of a microwave testbed for directionally modulated transmission. Directional modulation enables multiple communication and/or radar signals to be transmitted in multiple directions simultaneously using a single phased array aperture, helping to relieve spectral congestion. A two-element transmitter array is driven by a Xilinx ZCU208 Radio Frequency System on a Chip (RFSoC). Our testbed provides a foundation for developing a fully reconfigurable array transmitter for multi-user multiple-input multiple-output (MU-MIMO) radar and communications, which will incorporate in-situ measurement, reconfigurable matching circuitry, and fast tuning algorithms for frequency and directional selectivity. This testbed enables development and validation of reconfigurable techniques for adaptive spectral and spatial coexistence.

## I. INTRODUCTION

The congested wireless spectrum has led to coexistence concerns. Agile spectrum and spatial usage are needed, and both are possible because the spatial dimension can be harnessed for coexistence. Directional modulation (DM) that can simultaneously transmit multiple messages from the same aperture in multiple directions at the same frequency and time (Fig. 1). This paper presents a DM-capable transmitter testbed for developing reconfigurable hardware and MIMO arrays.

Babakhani demonstrates direction-dependent information transmission using near-field antenna modulation [1]. Daly describes how DM modulates the channel with optimized, constant-modulus element excitation weights [2]. Ansari overviews one-dimensional (angle only) and two-dimensional (angle and range) DM [3]. Snow describes the application of DM to radar and digital arrays [4, 5]. McCormick uses a single phased array for simultaneous radar and communications [6]. Huang describes how DM patterns can be used to direct beams without phase shifters using element switching [7]. Xie shows computationally efficient calculation of DM element excitations [8]. Here we demonstrate a DM testbed for microwave circuit and system developments, facilitating the consideration of reconfiguration on MIMO array transmissions.

The desired received signals in the far field (**r**), with received signals $r_n$ in directions $\theta_n$, are given in terms of the array matrix **H** and the weights of the different antenna excitations (**w**):

$$\begin{bmatrix} r_1 \\ \vdots \\ r_N \end{bmatrix}_{N \times 1} = \begin{bmatrix} \mathbf{h}_{11} & \cdots & \mathbf{h}_{1M} \\ \vdots & \ddots & \vdots \\ \mathbf{h}_{N1} & \cdots & \mathbf{h}_{NM} \end{bmatrix}_{N \times M} \begin{bmatrix} w_1 \\ \vdots \\ w_M \end{bmatrix}_{M \times 1} \quad (1)$$

Equation (1) can be solved for **w** with the Moore-Penrose pseudoinverse of **H** or other methods. We show the Channel State Information (CSI) approach for constructing **H** to demonstrate the baseline system, but traditional beam steering or Active Element Pattern (AEP) can alternatively be used.

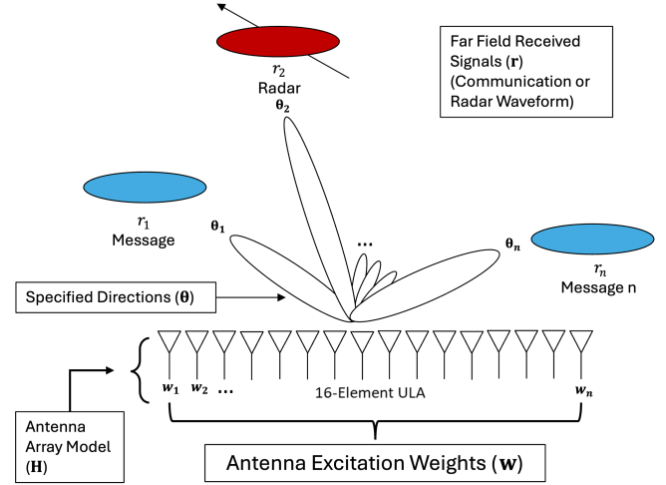

Fig 1. DM example scenario with a Uniform Linear Array (ULA) for MIMO, directional radar and communications

## II. DIRECTIONAL MODULATION TESTBED IMPLEMENTATION

Fig. 2 shows the design concept for a fully reconfigurable transmit array, consisting of three main subsystems: Reconfigurable Hardware, In-Situ Feedback, and Directional Modulation. Functionality planned for inclusion includes full reconfigurability with adaptive impedance tuning [9, 10] and in-situ measurement [11]. A Radio Frequency System-on-a-Chip (RFSoC) has been chosen for transmitter control because it has two transmit channels for the two transmitting elements and four receive channels to facilitate in-situ measurement for our current test array (two channels required per antenna).

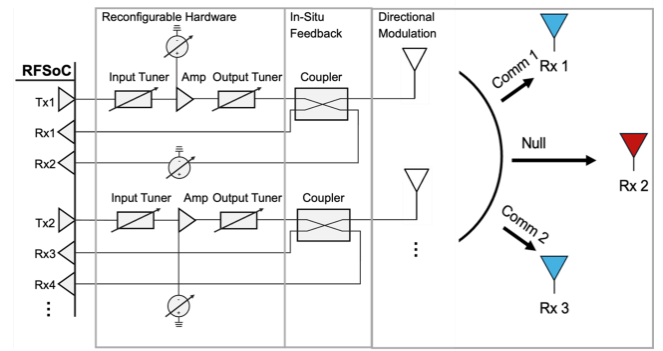

Fig 2. Planned fully reconfigurable transmit antenna array for DM, including reconfigurable input and output impedance matching networks, dynamic amplifier biasing, and in-situ measurement

The DM block of Fig. 2 has been implemented in this paper as the first step of constructing and validating this ultimate system.

Fig. 3 shows the testbed block diagram for this initial test, including a Xilinx ZCU208 RFSoC evaluation kit for independent phase and amplitude control of each transmit channel, a narrowband balun board to convert from the balanced line on the RFSoC to the unbalanced SMA connector, and a Mini-Circuits VBFZ-3590-S+ bandpass filter with a 3-4.3 GHz passband in each transmit channel. Fig. 4 shows a test setup photo, with approximately 4 feet between the transmit antenna and the two receive antennas (80° and 165°).

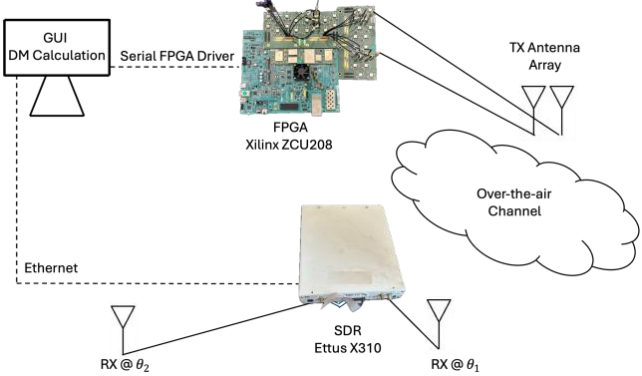

Fig 3. Directional modulation testbed block diagram, with transmit antenna inputs provided by the Xilinx ZCU208 RFSoC evaluation kit, and directional received signal measurements from the National Instruments Ettus X310 software-defined radio.

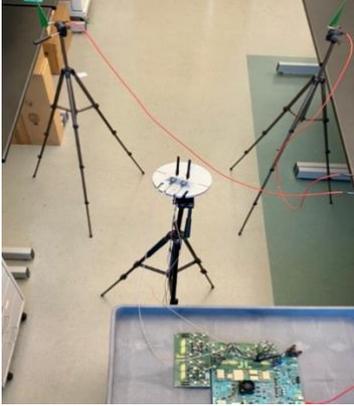

Fig 4. Measurement setup, with the RFSoC in the foreground connected to a turntable holding the two-element transmit array. Two receiving antennas act as multiple users in our MU-MIMO system. The antennas shown at the top of the image are placed on tripods about 4 feet from the transmit array at 80° to Rx 1 and 165° to Rx 2 (with respect to endfire on the right)

## III. Over-the-Air Calibration

The elements of $\mathbf{H}$ provide the phase and amplitude change between each transmitter element (column) and receiver (row):

$$\mathbf{H} = \begin{bmatrix} |T_{1,1}| & |T_{1,2}|e^{j\theta_1} \\ |T_{2,1}| & |T_{2,2}|e^{j\theta_2} \end{bmatrix}, \qquad (2)$$

where $T_{n,m}$ is the contribution to the signal received by receiver $n$ from the transmitter $m$, and $\theta_n$ is the phase difference between both transmitters as observed by receiver $n$. Four calibration transmissions can thus be used to construct the CSI and demonstrate its validity as follows:

(a) Operate the first transmitter at maximum amplitude, deactivating the second transmitter, to find $|T_{1,1}|$ and $|T_{2,1}|$.

(b) Operate the second transmitter at maximum amplitude, deactivating the first transmitter, to find $|T_{1,2}|$ and $|T_{2,2}|$.

(c) Operate both transmitters at maximum amplitude with no programmed phase difference. Measure the amplitude at each receiver. The measurement made at receiver $n$ has an amplitude given by

$$|T_{n,1} + T_{n,2}| = \sqrt{|T_{n,1}|^2 + |T_{n,2}|^2 + 2|T_{n,1}||T_{n,2}|\cos(\theta_n)}, \quad (3)$$

where $\theta_n$ is the phase difference between the two channels discerned at receiver $n$. Equation (3) is solved to find $|\theta_n|$:

$$|\theta_n| = \cos^{-1}\left(\frac{\left|T_{n,1} + T_{n,2}\right|^2 - \left|T_{n,1}\right|^2 - \left|T_{n,2}\right|^2}{2|T_{n,1}||T_{n,2}|}\right) \quad (4)$$

Because the cosine is an even function, another step is needed to determine whether $\theta_n$ is positive or negative.

(d) Operate both transmitters at maximum amplitude with a 90° programmed phase difference. Measure the amplitude at each receiver, denoted as $\left|T_{n,1} + T_{n,2}e^{j\frac{\pi}{2}}\right|$. For each $n$ ($n = 1, 2$), if $\theta_n \geq 0$, then

$$\left|T_{n,1} + T_{n,2}e^{j\frac{\pi}{2}}\right| = \left||T_{n,1}| + |T_{n,2}|e^{j(\frac{\pi}{2} + |\theta_n|)}\right| \quad (5a)$$

Otherwise, if $\theta_n \leq 0$, then

$$\left|T_{n,1} + T_{n,2}e^{j\frac{\pi}{2}}\right| = \left||T_{n,1}| + |T_{n,2}|e^{j(\frac{\pi}{2} - |\theta_n|)}\right| \quad (5b)$$

These two alternatives can be calculated by using the $\theta_n$ found in step (c) and finding which expression ((5a) or (5b)) best predicts the measured amplitude from step (d). This gives the sign of $\theta_n$.

## IV. Measurement Results

Experimental verification of DM at 4.2 GHz was performed using the testbed. The user controls and views results from the testbed using the MATLAB GUI shown in Fig. 5. In the Fig. 5 example, two messages were sent to different receivers. Following the launch of the MATLAB GUI and LabVIEW program, the user must first obtain $\mathbf{H}$ with the calibration procedure described in Section IV by clicking "Calibrate H".

Forward Error Correction (FEC), implemented using Low-Density Parity Check codes, can be toggled on and off using the "Enable FEC" switch. FEC was turned off for this example. The "Bits per Symbol" slider controls the DPSK constellation density. To begin transmission, the user presses the "Start Transmit" button. The weights are calculated based on $\mathbf{H}$, and then these weights are transmitted over the air from the two transmitters in the measurement setup. The calculated weights are shown in the plots on the top and bottom left of the GUI under the "Transmit Weights" header. The received phases at each receiver are plotted over time, and are decoded and displayed in the "1st Received Message" and "2nd Received Message" fields. In this example, the user has chosen to use 1 bit per symbol. The number of bit errors is also tracked and displayed per channel for simple viewing.

The Fig. 5 experiment shows the successful transmission of two different messages: "To satisfy some very young mathematician." (Message 1 to Receiver 1) and "It should be obvious." (Message 2 to Receiver 2). Alternatively, the user can also press the "Generate Message" button to generate random messages using the `why` function in MATLAB. If these messages are not of equal length, null characters are appended to the end of the shorter message to provide equal length. The received messages for both receivers match the transmitted messages in the Fig. 5 GUI, with no bit errors obtained out of 336 transmitted bits per channel. On the right side of the GUI, the plots of the received phases of the different receivers are shown. As expected for 1-bit DPSK, the measured phase transition is near either +90˚ or -90˚.

For further testing, 100 messages, each made up of 100 randomly generated ASCII characters, were transmitted on each channel without error correction and a DPSK constellation density of 1 bit per symbol, and the bit error rate (BER) on receive was evaluated. Table I summarizes these results. For 100 messages of 100 characters each, Channel 1 incurred 5.98% BER and Channel 2 incurred 7.67% BER, attributed to bit insertion errors. A bit insertion error occurs when the asynchronous phase detector incorrectly detects a non-existent symbol between actual transmitted symbols, as shown in an example in Fig. 6, where an uncompensated bit insertion error occurs at bit position 10. The false bit insertion causes a one-bit delay for the remaining actual bits, resulting in many of the subsequent message bits being decoded incorrectly. Bit insertion errors can be resolved by using a synchronized sampling system, which we plan to include in future work.

TABLE I: BIT ERROR RATE (BER) STATISTICS FOR 100 RANDOMLY GENERATED MESSAGES, EACH WITH A LENGTH OF 100 ASCII CHARACTERS (80,000 BITS TRANSMITTED ON EACH CHANNEL)

|  | Total Bit Errors | % Bit Error | Mean Bit Errors | STD Bit Errors |
|---|---|---|---|---|
| Channel 1 | 4780 | 5.98% | 47.80 | 114.37 |
| Channel 2 | 6132 | 7.67% | 61.32 | 132.12 |

To examine if bit insertion errors were the cause of the high bit error rate, a second experiment was conducted with shorter messages. After calibration, 1000 messages were transmitted, each composed of 10 randomly generated ASCII messages without error correction. The total number of transmitted bits in the multiple short messages of experiment 2 was equal to the total number of bits in the longer messages of experiment 1. A DPSK constellation density of 1 bit per symbol was used. The results, shown in Table II, indicate that the BER has been halved from the original, longer message scenario. With the reduction in bit insertion errors, most of the errors in this experiment are bit flip errors. Low-Density Parity Check error correction can correct bit flip errors if message length is known [12]. Fig. 7 shows bit flip errors at bit positions 13 and 15.

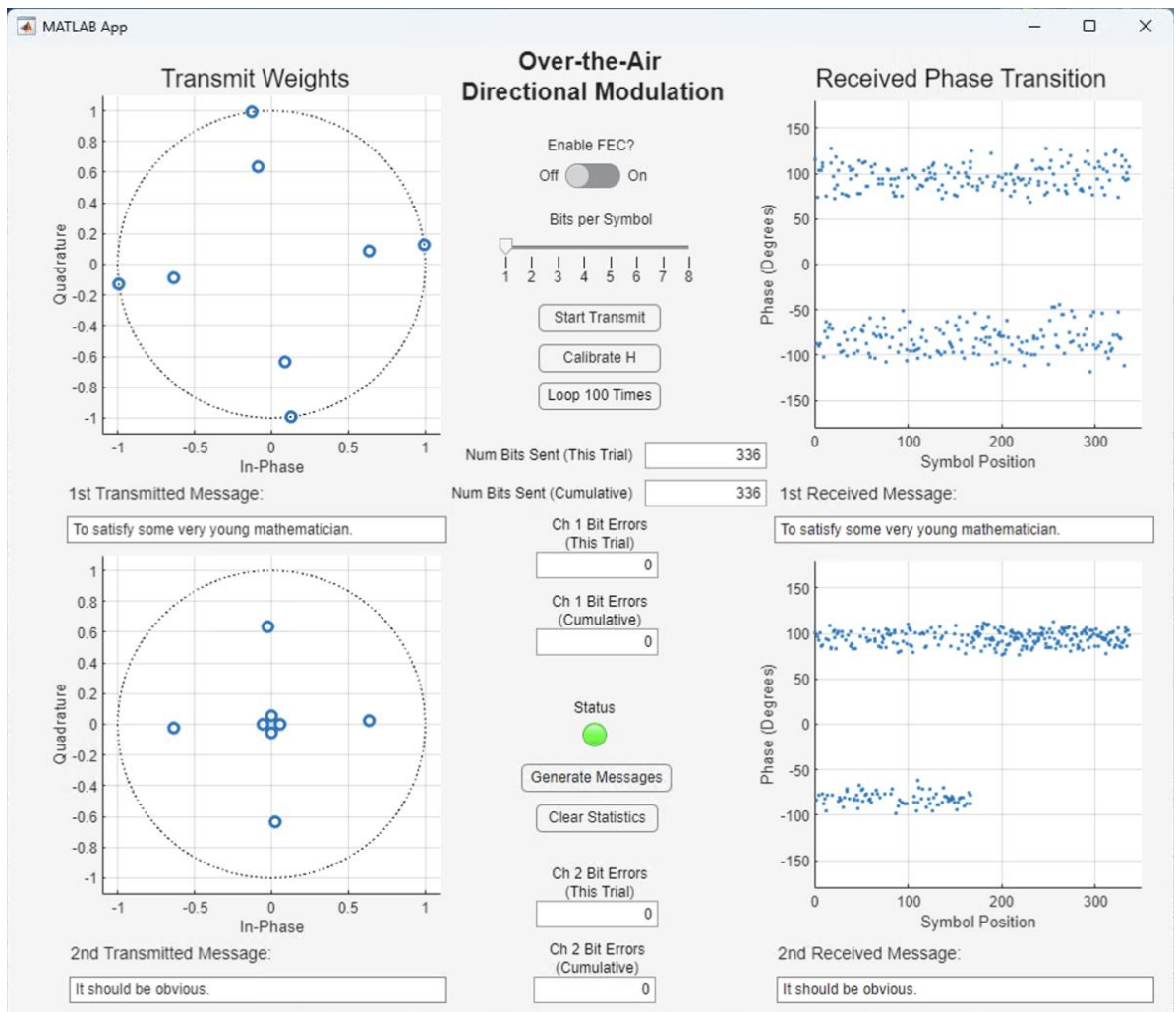

Fig 5. Graphical user interface mapping two messages to the two receiver used in the **H** matrix calibration. The two left plots show the in-phase and quadrature components of the transmit weights on the first antenna (top) and second antenna (bottom). The received phases are shown in the two right plots. The phases received at the first receiver are shown on the top, and the phases received at the second receiver are shown on the bottom.

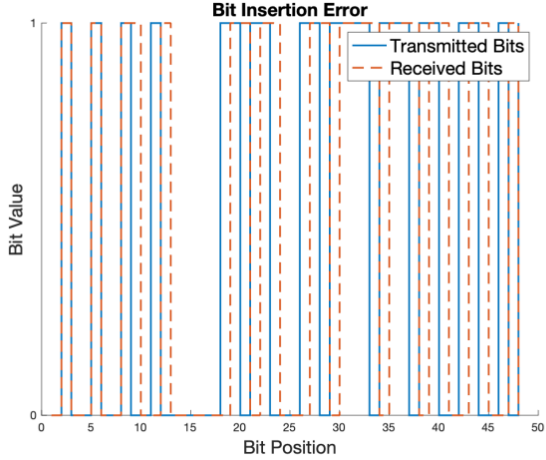

Fig 6. Transmitted and received bits (the first 48 bits of the 424-bit message) for the Table I experiment, with an Uncompensated Bit Insertion Error visible at bit position 10 and all subsequent bits shifted

TABLE II: BIT ERROR RATE (BER) STATISTICS FOR 1000 RANDOMLY GENERATED MESSAGES, EACH WITH A LENGTH OF 10 ASCII CHARACTERS (80,000 BITS TRANSMITTED ON EACH CHANNEL)

|  | Total Bit Errors | % Bit Error | Mean Bit Errors | STD Bit Errors |
|---|---|---|---|---|
| Channel 1 | 1665 | 2.08% | 1.67 | 7.49 |
| Channel 2 | 1901 | 2.38% | 1.90 | 7.77 |

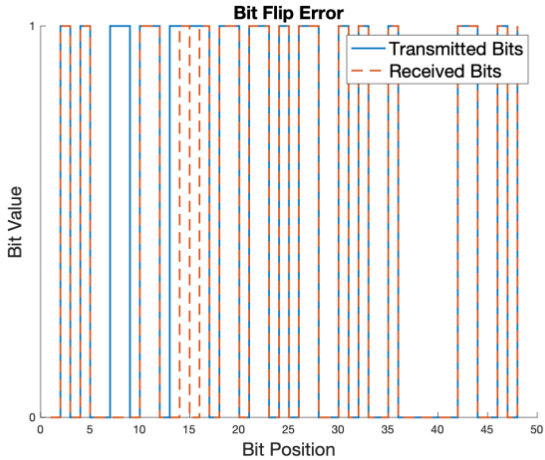

Fig 7. Transmitted and received bits (the first 48 bits of the 712-bit message) for the Table II experiment, with bit-flip errors visible at positions 13 and 15

## V. CONCLUSIONS

The implementation and validation of a testbed to implement and assess directionally and spectrally adaptive array technologies, such as for DM, has been detailed. DM, including calibration, has been demonstrated using a custom hardware implementation. This foundation provides the capability to demonstrate and innovate the impacts of spatial modulation (such as DM), real-time impedance tuning, and in-situ measurement in an array for spectrally and spatially agile MIMO systems.